\documentclass[twocolumn,showpacs,preprintnumbers,amsmath,amssymb,superscriptaddress]{revtex4}
\usepackage{graphicx}

\newcommand{\beq}{\begin{equation}}
\newcommand{\eeq}{\end{equation}}
\newcommand{\beqn}{\begin{eqnarray}}
\newcommand{\eeqn}{\end{eqnarray}}

\begin{document}

\title{Time-reversal symmetry breaking phase in 
the
Hubbard model: a VCA study} 

\author{Xiancong Lu}
\affiliation{Department of Physics and Institute of Theoretical Physics and Astrophysics, Xiamen University, Xiamen 361005, China}
\affiliation{Institute of Theoretical and Computational Physics, Graz
  University of Technology, A-8010 Graz, Austria}

\author{Liviu Chioncel}
\affiliation{Augsburg Center for Innovative Technologies, University of Augsburg, D-86135 Augsburg, Germany}
\affiliation{Theoretical Physics III, Center for Electronic Correlations and Magnetism, Institute of Physics,
University of Augsburg, D-86135 Augsburg, Germany}

\author{Enrico Arrigoni}
\affiliation{Institute of Theoretical and Computational Physics, Graz
  University of Technology, A-8010 Graz, Austria}

\begin{abstract}
We study the stability of the time-reversal symmetry breaking 
staggered flux
phase
of a single band 
Hubbard model, 
within the Variational Cluster Approach (VCA).
For intermediate and small values of the interaction $U$, we
find 
metastable 
solutions for the staggered flux phase,
with a 
maximum current per bond 
at 
$U\approx 3.2$. 
However, 
allowing for 
antiferromagnetic and superconducting long-range order
 it turns out that in the region at and 
close to half filling
the antiferromagnetic phase is 
the  most favorable energetically.
The effect of nearest-neighbour interaction is also considered. 
Our results show
 that a negative nearest-neighbour
interaction and finite doping 
favors the stability of the staggered-flux phase.
We also present preliminary results for the three-band Hubbard model
obtained with a restricted set of variational parameters.
For this case, 
no spontaneous time-reversal
symmetry breaking phase is found in our calculations.

\end{abstract}

\pacs{71.10.Fd, 71.27.+a, 71.10.Hf, 74.20.-z}
\maketitle

\section{Introduction}

The Hubbard Hamiltonian  is perhaps the simplest possible model  
to describe electronic interactions in
solids. 
It is however,
difficult to solve and even after several decades of research there are still many 
open questions about its basic features. In particular many competing phases have 
been proposed as candidates for the ground state in special parameter regions. 
The staggered flux phase~\cite{AffleckPRB88,MarstonPRB89}, known also as "orbital
antiferromagnet" \cite{HalperinSSP68}, was proposed to describe the
ground state of Heisenberg Antiferromagnetism. Many properties of the
staggered flux phase have been discussed in the past.
 However, less 
numerical evidence exists that support the existence of this phase in the
Hubbard or t-J model~\cite{ko.li.88.sm,le.po.89,zhan.90,sc.wh.01,MacridinPRB04,LeungPRB00,StanescuPRB01}.  
The phenomena of pseudogap in cuprate superconductor contributed to the  
increased interest in the time-reversal symmetry broken phase. 
Chakravarty \textit{et al.}~\cite{ch.la.01} argued that %the strange phenomenology in 
the pseudogap region is due to the competition between $d$-wave superconductivity
and $d$-density wave (DDW) state \cite{NayakPRB00,SchulzPRB89}, which
actually is the staggered flux state breaking both time-reversal and
translational symmetry. Based on the mean-field analysis of three-band
Hubbard model, Varma 
showed
 that a time-reversal symmetry breaking
phase, the ``circulating current phase''
\cite{varm.97.nf,varm.99,si.va.02,varm.06}, is stable in some
parameter regions. The orbital current of this phase is
circulating along the O-Cu-O plaqette in each cell and thus,
in contrast to DDW, 
 translational symmetry 
is preserved. Motivated by this proposal, several theoretical and
experimental groups
have tried
to find the signatures of orbital current in the CuO planes
\cite{gr.th.07,th.gr.08,fa.si.06,ma.ac.08}. However, 
no agreements have been obtained 
so far.
Recent polarized neutron
diffraction experiment shows 
that the spontaneous current occurs in loops involving apical oxygen orbitals
\cite{li.ba.08}. This picture is consistent with 
Variational Monte Carlo (VMC) calculations
 \cite{we.la.09}, although
the authors observe that the computed current 
of three-band Hubbard model 
is smaller, for larger  
system sizes, therefore computations on larger system are required.

There 
have been
 a lot of efforts to search 
for
the staggered flux phase in the
two-leg ladder, which is easier to study and can shed some light on
the full two dimension lattices. By using the highly accurate
density-matrix renormalization group technique, 
Marston \textit{et. al.}~\cite{ma.fj.02} and 
Schollw\"ock \textit{et al.}~\cite{sc.ch.03} found the evidences 
for the existence of staggered flux phase at and away from half 
filling, respectively. 
It has been shown
that complex interactions are 
needed to stabilize this phase
\cite{ma.fj.02,sc.ch.03}. 
Analytical studies using
bosonization/renormalization group (RG) method also found 
 stable
regions of this phase for weak interactions
\cite{schu96,or.gi.97,fj.ma.02,wu.li.03}.
In the present work we 
adopt the
Variational Cluster Approach (VCA) to study the time-reversal 
symmetry breaking phase. VCA allows to evaluate the 
single-particle~\cite{da.ai.04} and two-particle~\cite{br.ar.10} 
spectral functions, and, to include symmetry breaking 
phases~\cite{da.ai.04,se.la.05,ai.ar.05}, such as Antiferromagnetism
and 
Superconductivity, therefore it can be easily extended to include 
time-reversal symmetry breaking phase. 

The paper is organized as follows. In Sec. \ref{vca}, we 
present the Hubbard Hamiltonian in a form that includes the single  particle
parameters that are relevant for the time-reversal symmetry breaking.
We review briefly the VCA method and 
introduce the coupling fields associated with
the time-reversal symmetry breaking,
so called
 "Weiss" fields.
These are the hopping 
$\Delta t$ and 
its  phase  $\Delta\phi$, together with the current
per bond which arises as a natural order parameter. 
The search for the spontaneous time-reversal breaking state within the interacting 
Hubbard model is presented in Sec. \ref{fl2dh}. 
The following sections
Secs. \ref{fl2dha} and \ref{fl2dhb} discuss the stability of staggered flux phase at 
and away from half-filling for the single band %non-magnetic %(paramagnetic) interacting 
Hubbard model. 
Our results 
indicate
 that 
the
stable 
ground state for the interacting Hubbard model at half
filling is the antiferromagnetic state. 
Away 
from half-filling,
 the competition with the superconducting phase is
also investigated. Sec. \ref{nnU} discusses the effect of
nearest-neighbour interaction on the staggered-flux phase. In
Sec. \ref{3band}, we extend the procedure described in Sec. \ref{vca}
to the three band Hubbard model and search 
for a
 spontaneous time-reversal
symmetry breaking phase in 
this model. 
Finally, we draw our conclusions in
Sec. \ref{con}.

\section{Hamiltonian and Variational Cluster Approach}
\label{vca}
We consider the following 2D Hubbard model 
on a square lattice:
\begin{equation}\label{Hubb}
H = -   \sum_{<ij>\sigma} ( 
\widetilde{t}_{ij}
%%{t}_{ij}
 c_{i\sigma}^\dag c_{j\sigma} + h.c. )
    + U \sum_in_{i\uparrow}n_{i\downarrow}
\end{equation}
where $c^\dag_{i\sigma}$($c_{i\sigma}$) creates (destroys) an electron
on site $i$ with spin $\sigma$,
$n_{i\sigma}=c^\dag_{i\sigma}c_{i\sigma}$ is the particle number
operator, $\langle ij\rangle$ denotes 
 nearest neighbor sites,
and $U$ is the local Coulomb
interaction. 
$\widetilde{t}_{ij}= t e^{i\phi_{ij}}$
 is the hopping-matrix element,
where $t$ is a real number and $\phi_{ij}$ is the phase factor on the
bond $\langle ij\rangle$.
In the following, we take $t=1$ to be our unit of energy.
 If all $\phi_{ij}=0$, Eq. (\ref{Hubb}) is the Hamiltonian
of the standard Hubbard model with real hopping matrix. For
$\phi_{ij} \ne 0$, Eq. (\ref{Hubb}) describes electrons hopping on 
lattices subjected to a magnetic field perpendicular to the 
plane of lattices
provided the circulation of $\phi_{ij}$ is nonzero.
For $\phi_{ij}=0$ the Hamiltonian is symmetric under time reversal.

The Variational Cluster Approach (VCA) \cite{po.ai.03,da.ai.04}
employed in this paper 
can be seen as an extension of
Cluster Perturbation
Theory (CPT) \cite{gr.va.93,se.pe.00,ov.sa.89} based on the
Self-energy Functional Approach (SFA) \cite{pott.03,pott.03.se}. 
One shortcoming of CPT method is that it can not study the
spontaneous symmetry-breaking phase. As an improvement over it, the
VCA adds additional ``Weiss'' field (describing a particular ordered state)
to the cluster Hamiltonian, and
optimizes 
the self-energy of the reference cluster by variational
principle. The value for the variational parameter is determined by
searching the saddle point of SFA grand potential,
\begin{eqnarray}
\label{omega}
\Omega=\Omega'+\mathrm{Tr}\ln(\mathbf{G}_0^{-1}-\mathbf{\Sigma})^{-1}-\mathrm{Tr}\ln\mathbf{G}'
\end{eqnarray}
where $\mathbf{G}_0$ is the non-interacting Green's function,
$\Omega'$, $\mathbf{\Sigma}$, and $\mathbf{G}'$ are the
grand-canonical potential, self energy, and Green's function of the
reference cluster, respectively.
The system is
in the ordered state if the corresponding ``Weiss'' field is nonzero
at the saddle point of grand potential $\Omega$. %VCA has been

For an initially time-reversal symmetric Hamiltonian ($\phi_{ij}=0$), 
spontaneous symmetry breaking can occur under some circumstances.
Within VCA this can be studied by introducing in the reference system
an appropriate
``Weiss'' field \cite{da.ai.04}. 
The ``Weiss'' field ($ H_{TR}'$) that breaks 
the time-reversal symmetry of the cluster Hamiltonian 
can be written in the following form:
\begin{eqnarray}\label{htr}
 H_{TR}'= \sum_{<ij>\sigma}(\Delta t\; e^{i\Delta\phi_{ij}}
          c_{i\sigma}^\dag c_{j\sigma} + h.c. )
\end{eqnarray}
Here, $\Delta t$ is the amplitude of hopping acting as the
strength of the ``Weiss'' field, and  $\Delta\phi_{ij}$ is 
the phase factor on bond $\langle ij \rangle$ which 
determines the
phase configuration of time-reversal symmetry-breaking phase. Both
$\Delta t$ and $\Delta\phi_{ij}$ should be treated as variational
parameters when searching the saddle points of the SFA grand potential
$\Omega$ in Eq. (\ref{omega}). 
A time-reversal symmetry-breaking phase 
is characterized by a saddle point of $\Omega$ with nonzero values
of both $\Delta t$ and $\Delta\phi_{ij}$.

\begin{figure}[h]
\includegraphics[width=0.65\columnwidth]{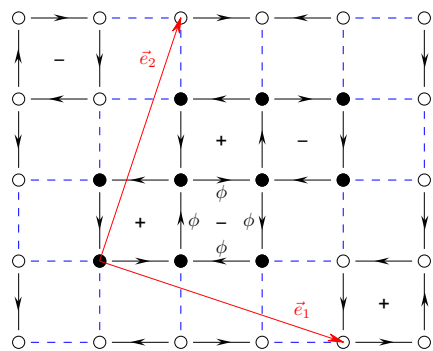}
\caption{(Color online). Staggered flux 
pattern in a 2D square lattice. $\phi$ is the phase
   on each bond in the direction indicated by the arrow.
 The total magnetic flux through a plaquette is
  $4\phi$. $+$ and $-$ denote the directions of the corresponding magnetic
  field. The black dots denote the 10-site reference cluster we use in
  our calculations. The vectors $\vec{e}_1$ and $\vec{e}_2$ are the
  translation vectors of VCA. }
\label{stagflux}
\end{figure}

The natural order parameter for the time-reversal symmetry-breaking phase
is the current. For a typical bond $\langle lm \rangle$, 
the current is defined as
\begin{eqnarray}\label{curr}
J_{lm}= \sum_\sigma 
     i \widetilde{t}( < c_{l\sigma}^\dag c_{m\sigma} >
                    - < c_{m\sigma}^\dag c_{l\sigma} > )
\end{eqnarray}
It is obvious that $J_{lm}$ is nonzero only if the time-reversal
symmetry is broken. 
In the staggered flux state the phase displays a pattern like the one
illustrated in  Fig.~\ref{stagflux}. 
If no other symmetries are broken one expects the absolute value of
the phase to be the same on all bonds. We thus take
$|\Delta\phi_{ij}|=\Delta\phi$ to be the same on all bonds~\cite{footnote}.

Before proceeding to search for the spontaneous time-reversal symmetry
breaking phase,
 we first  test our  code by introducing an external
 staggered flux phase on a square lattice induced by an
external magnetic field.
The dynamics of  two dimensional lattice electrons coupled to an external
magnetic field is described by the
 tight-binding
Hamiltonian
in Eq. (\ref{Hubb}), with 
the circulation of $\phi_{ij}$ 
on a given path being proportional to the magnetic 
flux through the enclosed surface~\cite{hofs.76}.

For a staggered flux pattern and
interaction $U=0$, the model can be
 solved analytically and the energy spectrum obtained.
The ground state $E_G$ and current $J$ can be
computed by summing all the energies up to the Fermi energy and by 
the
Hellmann-Feynman theorem. 
We carry out our test  in the non-interacting limit for   a staggered magnetic field
 with a phase pattern as shown in Fig. \ref{stagflux}. 
All the quantities evaluated:
spectral function, $E_G(\phi)$, and $J$ using 
our VCA code
agree precisely with the analytical 
results.

For the case of interacting electrons 
 we
additionally checked the validity 
of 
the 
implementation by verifying thermodynamic consistency.
Specifically concerning the  current,
 its expectation values 
 evaluated via the
spectral function and via the Hellmann-Feynmann theorem 
turn out to be identical.
Although this is obviously expected, it in practice non trivial since,
as discussed in \cite{ai.ar.06}, this is only guaranteed if one uses
the corresponding hoppings as variational parameters.
By separating the hopping term in Hamiltonian (\ref{Hubb}) into real
and imaginary parts: $\alpha (c_{i\sigma}^\dag c_{j\sigma}+c_{j\sigma}^\dag c_{i\sigma})$ and 
$i \beta (c_{i\sigma}^\dag c_{j\sigma}-c_{j\sigma}^\dag c_{i\sigma})$, with $\alpha=-t\cos(\phi)$ and
$\beta=-t\sin(\phi)$, 
we confirm that
treating $\beta$ as variational parameter the thermodynamic
consistency condition discussed above is satisfied.

\section{Spontaneous staggered flux phase in the 2D Hubbard model}
\label{fl2dh}

\begin{figure}[h]
\includegraphics[width=0.90\columnwidth]{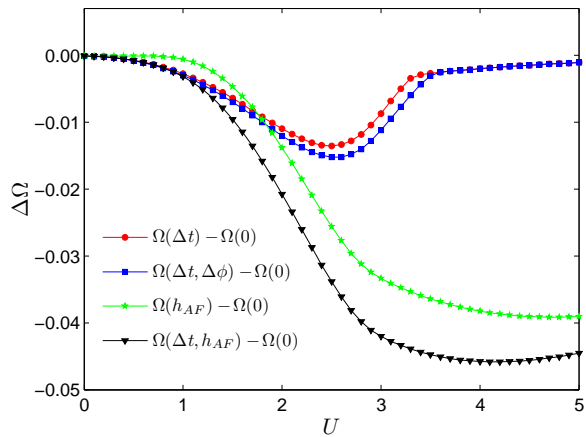}
\caption{(Color online). Difference between the grand potentials computed
with variational parameters, such as the hopping $\Delta t$, 
phase factor $\Delta\phi$, and staggered magnetic field $h_{AF}$,
with respect to the grand potential $\Omega(0)$ when no variational
parameters are considered. The calculations are performed at half filling.}
\label{deltaOmega}
\end{figure}

In this section, we study the stability of the spontaneous
staggered flux phase in the Hubbard model with \textit{real}
hopping matrix (i.e., $\phi_{ij}=0$ in Eq. (\ref{Hubb})).
 The
situation is different from what we discuss in 
Section \ref{vca}
where the electrons are under an external staggered magnetic field.
In order  to allow for the time-reversal symmetry-breaking order, we introduce
a Weiss field $H_{TR}'$ (cf. Eq. (\ref{htr})) in the reference system.
 For the staggered flux phase,
the absolute value of the phase  $|\Delta\phi_{ij}|$ is chosen to be equal for
all bonds (i.e, $\Delta\phi$) and the staggering configuration is
shown together with the  VCA reference cluster in Fig. \ref{stagflux}. 
We stress that it is important to choose a reference cluster with 
even number of plaquettes, so that the net ``magnetic field'' in 
the cluster is zero.

\subsection{Staggered flux phase in the Hubbard model at half-filling}
\label{fl2dha}

In Fig. \ref{deltaOmega} we plot the difference between the values of the 
grand potential obtained when variational parameters are included relative
to the cases when no variational parameters are considered. The results 
presented are for the case of half-filling and different values of U. 
In the small $U$ region (i.e., from 0 to 3.6), the values of the grand potential
$\Omega(\Delta t,\Delta\phi)$ with both $\Delta t$ and $\Delta\phi$ taken as
variational parameters are \textit{always} smaller than the grand potential
$\Omega(\Delta t)$ with a single variational parameter $\Delta t$, the latter being
also smaller than the grand potential obtained in a calculation with no 
variational parameters $\Omega(0)$. Accordingly, the staggered flux phase 
is lower in energy 
than
the correlated paramagnetic (non-magnetic) state of the 
tight-binding Hamiltonian at half-filling.
At $U\approx 3.6$ the optimal $\Delta \phi$ vanishes indicating a
phase transition to a phase without time-reversal symmetry breaking.
This suggests that the staggered flux phase is 
favored by small values of
 $U$.
 This situation was not 
considered within  Ref. \cite{AffleckPRB88}, which predicts that the 
staggered flux phase is stable at large $U$ limit, a consequence of the 
mean-field approach.
At half filling one should expect that
 the ordered 
Antiferromagnetic phase 
also plays an important role in the Hubbard model. 
 In order to 
compare the stability of the flux phase with respect to the Antiferromagnetic 
long-range order we 
included a Neel Antiferromagnetic  Weiss field
\begin{equation}
H'_{AF}=h_{AF}\sum_i(n_{i\uparrow}-n_{i\downarrow})e^{i\mathbf{Q}\mathbf{R}_i}
\end{equation}
into the reference system.
Here,
 $h_{AF}$ is the strength of a staggered magnetic field and
$\mathbf{Q}=(\pi,\pi)$ is the Antiferromagnetic wave vector.
The calculations within the long-range ordered state %Fig. \ref{deltaOmega} 
shows that $\Omega(\Delta t,\Delta\phi)$ 
is larger than 
$\Omega(h_{AF})$ for values of $U>1.9$, 
and is lower than $\Omega(h_{AF})$ when $U<1.9$. 
But for a complete comparison, $\Omega(\Delta t,\Delta\phi)$ has to be compared with  
$\Omega(\Delta t, h_{AF})$ when both $\Delta t$ and $h_{AF}$ are
treated as variational parameters
in the long-range ordered state. As shown in Fig. \ref{deltaOmega}, 
in this case
$\Omega(\Delta t,\Delta\phi)$ is \textit{always} larger 
than $\Omega(\Delta t, h_{AF})$ for the whole range of $U$.

Although 
our results indicate that 
the Antiferromagnetic state is more stable than
the staggered flux phase for the half-filled 
 2D Hubbard model,
the latter can be still considered as a metastable phase whose
fluctuations affect the physics of the system, or which may become
stable whenever Antiferromagnetic long-range order is suppressed by some other
mechanism.

\begin{figure}[h]
\includegraphics[width=0.98\columnwidth]{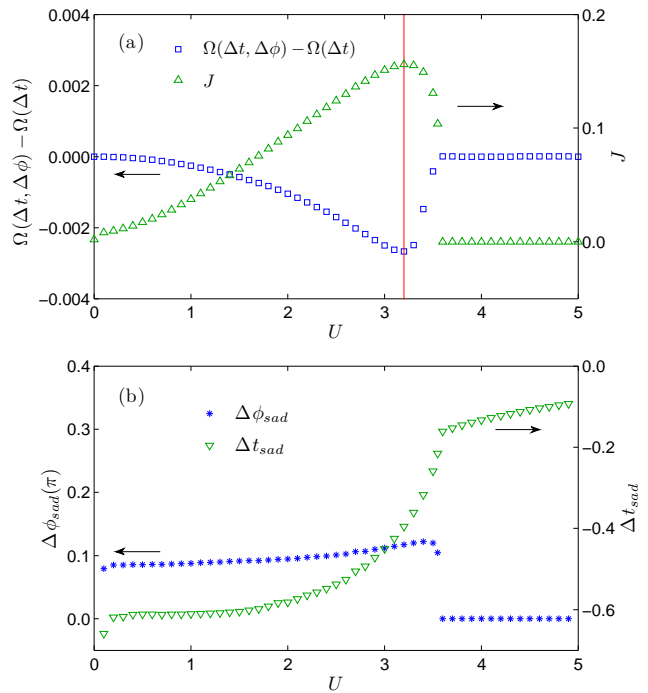}
\caption{(Color online). (a) The grand potential difference between
  $\Omega(\Delta t,\Delta\phi)$ and $\Omega(\Delta t)$ (denoted by
  $\square$ with labeling on the left) and the current $J$
  corresponding to $\Omega(\Delta t,\Delta\phi)$ (denoted by
  $\vartriangle$ with labeling on the right) as a function of
  interaction $U$. (b) The value of variational parameters at the
  saddle points of $\Omega(\Delta t,\Delta\phi)$, $\Delta\phi_{sad}$
  ($\ast$ with labeling on the left and in unit of $\pi$) and $\Delta
  t_{sad}$ ($\triangledown$ with labeling on the right), as a function
  of interaction $U$. The calculations are performed at half filling.}
\label{phimin}
\end{figure}

In order to study the metastable staggered flux state in more detail, we plot in
Fig. \ref{phimin}a the difference in grand potentials  
$\Omega(\Delta t,\Delta\phi) - \Omega(\Delta t)$ (symbol -- $\square$, 
with labeling on the left) and the corresponding current $J$ as a function 
of interaction $U$ at half filling. 
The difference between the values of 
the grand potential increases with U, reaches a maximum for the value 
of $U \approx 3.2$, and then 
decreases 
to zero for  $U \approx 3.6$. 
The largest difference between $\Omega(\Delta t,\Delta\phi)$ and 
$\Omega(\Delta t)$ is $0.0027$ which 
corresponds to about $1$ meV for $t\approx 500 meV$.
The current per bond $J$ (obtained in the staggered flux phase)
is computed 
by
treating both $\Delta t$ and $\Delta \phi$ as variational 
parameters. The computed values are shown in Fig. \ref{phimin}a with 
labeling on the right. One sees 
an opposite
behaviour 
of the current as a function of U
with respect to $\Omega(\Delta t,\Delta\phi) - \Omega(\Delta t)$.

Fig. \ref{phimin}b shows the values of variational parameters $\Delta
\phi_{sad}$ and $\Delta t_{sad}$  at the saddle points
 as a function of interaction $U$ at
half filling. One can see that the $\Delta\phi_{sad}$ increases as $U$
is increasing in the small $U$ region. After reaching its maximum
around $U \approx 3.5$, where both the difference $\Omega(\Delta
t,\Delta\phi) - \Omega(\Delta t)$ and the current $J$ become zero,
$\Delta\phi_{sad}$ drops sharply to zero. Note that, in the regions of
$U<0.7$ and $U \approx 3.5$, the difference between $\Omega(\Delta
t,\Delta\phi)$ and $\Omega(\Delta t)$ is very small, in the same order
as the accuracy of our calculation ($1.0\times 10^{-4}$). As in
Fig. \ref{phimin}b, the absolute value of variational parameter
$\Delta t_{sad}$ is decreasing as $U$ increases. The slope of $\Delta
t_{sad}$ curve changes abruptly at the place where the
value of $\Delta\phi_{sad}$ becomes zero.

\subsection{Staggered flux phase in the Hubbard model away from half filling}
\label{fl2dhb}
In this section, we study the stability of staggered flux phase when
the system is away from half filling. We consider the competition between 
the staggered flux phase, superconductivity, and anti-ferromagnetism. 
Away from half-filling, the shift in 
the chemical potential $\Delta\mu$ must be treated as variational parameter 
as well in order to satisfy the condition of thermodynamic consistency
\cite{ai.ar.05}. 
The
d-wave Superconducting phases~\cite{da.ai.04,se.la.05} 
can be described by 
introducing  the corresponding Weiss field $H'_{SC}$: 
\begin{equation}
  H'_{SC}=h_{SC}\sum_{ij}\frac{\Delta_{ij}}{2}(c_{i\uparrow}c_{j\downarrow}+H.c.),
  \label{hsc}
\end{equation}
where $h_{SC}$ is the strengths of the  nearest-neighbor d-wave pairing field 
and $\Delta_{ij}$ is the d-wave form factor.

\begin{figure}[h]
\includegraphics[width=0.88\columnwidth]{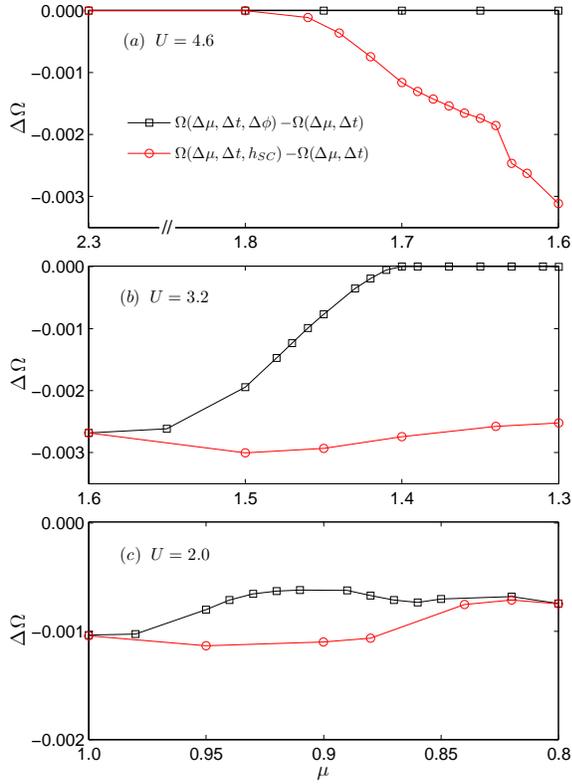}
\caption{(Color online). The grand potential of superconductor phase
$\Omega(\Delta\mu, \Delta t, h_{SC})$ and staggered-flux phase
$\Omega(\Delta\mu, \Delta t, \Delta\phi)$, with respect to
$\Omega(\Delta\mu, \Delta t)$, as a function of chemical potential
$\mu$ for various interactions $U$. $\Delta\mu$, $\Delta t$,
 $\Delta\phi$, and $h_{SC}$  %treating the shift of chemical
are variational parameters
.}
\label{delta_omega_mu}
\end{figure}

Let us start by comparing 
the superconducting and staggered flux states. 
In Fig. \ref{delta_omega_mu} we compare the grand
potential of the superconducting phase $\Omega(\Delta\mu, \Delta t,
h_{SC})$ with that of the staggered flux phase $\Omega(\Delta\mu, \Delta
t, \Delta\phi)$ for different values of the interaction parameter $U$. 
One can see that the values of the grand potential in the superconducting state
$\Omega(\Delta\mu, \Delta t, h_{SC})$ and in the 
staggered flux
phase $\Omega(\Delta\mu,\Delta t, \Delta\phi)$
are equal at half filling. This can be easily understood  because
the d-wave pairing field $H_{SC}'$ in Eq. (\ref{hsc}) is connected
to the staggered-flux field $H_{TR}'$ in Eq. (\ref{htr}) by a
particle-hole transformation at half-filling \cite{Lee06,al.zo.88}.
When away from half-filling, $\Omega(\Delta\mu, \Delta t, h_{SC})$ is
always \textit{smaller} than $\Omega(\Delta\mu, \Delta t, \Delta\phi)$
for all interactions, implying that 
the superconducting state is more stable than the staggered flux phase 
in this region.  
For the large $U$ case
($U \gtrsim 4.0$), we can not find a VCA saddle point corresponding to
the
staggered flux phase 
neither  at nor away
 from half filling. Here, the value of the
variational parameter $\Delta\phi$ is always zero 
as seen in 
Fig. \ref{delta_omega_mu}(a) for $U=4.6$.
Fig. \ref{delta_omega_mu}(a) shows that the values
of the grand potential $\Omega(\Delta\mu, \Delta t, h_{SC})$ are smaller than 
that of $\Omega(\Delta\mu, \Delta t)$ and the difference between them is 
increasing as the chemical potential $\mu$ decreases. For smaller values of $U$ 
a
 metastable VCA solution for staggered flux phase can be obtained, that is,  
$\Omega(\Delta\mu,\Delta t, \Delta\phi)$ is smaller than $\Omega(\Delta\mu, \Delta t)$
with a finite value of variational parameter $\Delta\phi$. For the
intermediate $U$ in Fig. \ref{delta_omega_mu}(b), the staggered flux phase
disappears at some value of $\mu$. Finally, for small $U$ the metastable VCA 
solution for the staggered flux phase always exists for the whole range 
of chemical potential as seen in Fig. \ref{delta_omega_mu}(c). 
Note that the plots
in Fig. \ref{delta_omega_mu}
are 
shown as a function of chemical potential, which corresponds to a
doping
in the range of 0 $\sim$ 0.1.

\begin{figure}
\includegraphics[width=0.90\columnwidth]{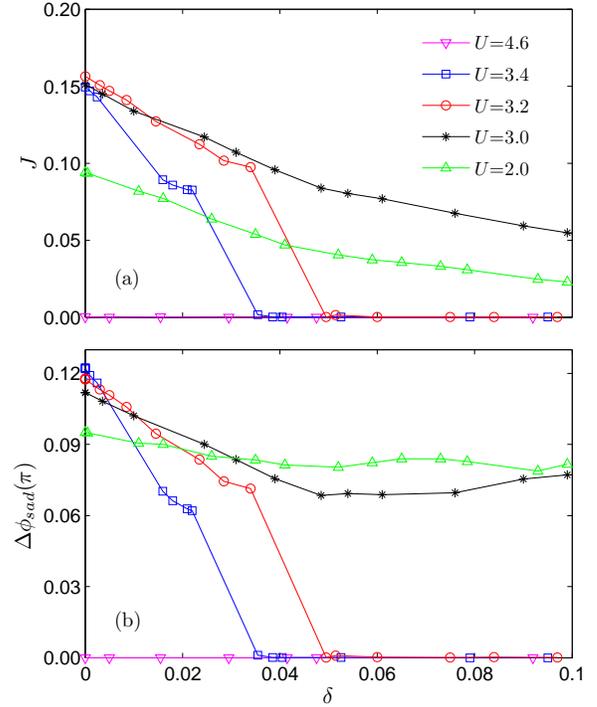}
\caption{(Color online). Current $J$ (a) and the value of
  variational phase factor $\Delta\phi_{sad}$ (b) at the saddle points
  of $\Omega(\Delta\mu,\Delta t,\Delta\phi)$ as a function of doping
  $\delta$ for various interactions $U$. 
}
\label{cur_phi_dop}
\end{figure}

In Fig. \ref{cur_phi_dop}, the current $J$ (a) and the value of
variational phase factor $\Delta\phi_{sad}$ (b) at the saddle points
of 
the staggered flux phase 
are plotted as a function of doping $\delta$
for various interactions $U$. One can see that both $J$
and $\Delta\phi_{sad}$ are zero for large 
$U\gtrsim 4.0$,
indicating no VCA solution for staggered flux phase in this
region. For intermediate $U$, $J$ and $\Delta\phi_{sad}$ decrease as the
doping $\delta$ is increasing and drop to zero at some 
value of
doping (e.g.,
$\delta=0.05$ for $U=3.2$). For small $U$ (e.g., $U=2.0$), it seems
like that $\Delta\phi_{sad}$ keeps as a finite value (around 0.8$\pi$)
and does not drop to zero. However, the staggered flux phase is 
less  stable than
the superconducting phase, as shown in Fig. \ref{delta_omega_mu}.

\begin{figure}[h]
\includegraphics[width=0.90\columnwidth]{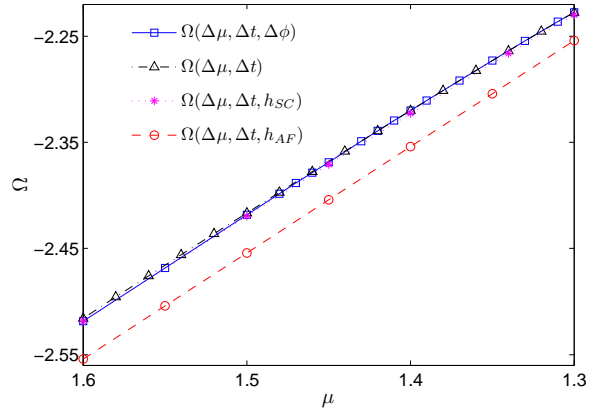}
\caption{(Color online). Grand potential $\Omega$ with different
variational parameters ($\Delta\mu$, $\Delta\phi$, $h_{SC}$, and $h_{AF}$) as a function 
of the chemical potential $\mu$ for $U=3.2$. %The $\Delta\mu$,
}
\label{omega_mu}
\end{figure}

Including the antiferromagnetic phase, we choose to look at the results for 
the values of $U=3.2$, where the difference between $\Omega(\Delta t,\Delta\phi)$ 
and $\Omega(\Delta t)$ is maximal at half filling (see Fig. \ref{phimin}a). 
In Fig. \ref{omega_mu} we show the grand potential $\Omega$ with different 
variational parameters as a function of chemical potential $\mu$. %at a fixed $U=3.2$. 
It is clear that $\Omega(\Delta\mu, \Delta t, h_{AF})$ is much smaller
than both $\Omega(\Delta\mu, \Delta t, h_{SC})$ and $\Omega(\Delta\mu, \Delta t,
\Delta\phi)$. Thus, the antiferromagnetic phase is the most stable
state 
in this region of chemical potential. For even smaller chemical
potential (or larger doping),
the antiferromagnetic phase 
becomes   unstable
 towards the superconducting 
phase.

In summary, we find that the staggered flux phase
is less stable than the superconducting state for all interactions
and chemical potentials considered.
 For values of doping not very far away from 
half filling, the antiferromagentic phase would be the most stable phase.

\subsection{Staggered flux phase in the Hubbard model with
  nearest-neighbor interaction}
\label{nnU}

\begin{figure}[h]
\includegraphics[width=0.90\columnwidth]{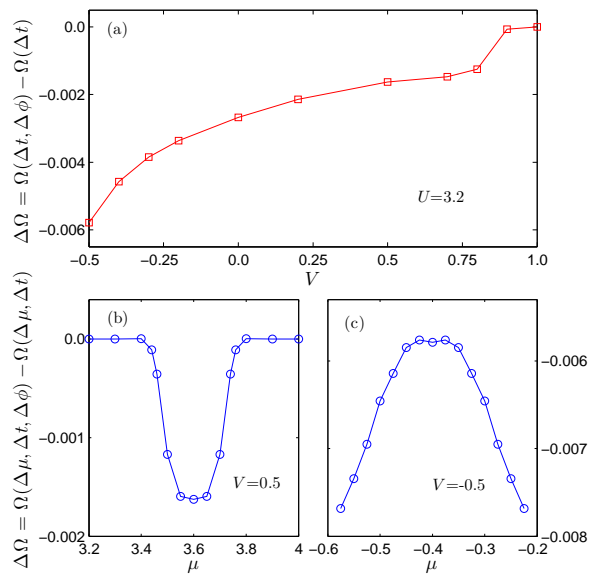}
\caption{(Color online). (a) The grand potential difference
  $\Omega(\Delta t, \Delta\phi)-\Omega(\Delta t)$ as a function of
  nearest-neighbor interaction $V$ at half-filling for a fixed on-site
  interaction $U=3.2$. (b) and (c) show the grand potential difference
  $\Omega(\Delta\mu,\Delta t,\Delta\phi)-\Omega(\Delta\mu,\Delta t)$
  as a function of chemical potential $\mu$ for nearest-neighbor
  interaction $V=0.5$ and $V=-0.5$ respectively. The on-site
  interaction is also fixed at $U=3.2$.  }
\label{delta_v}
\end{figure}

In this section, we discuss the effect of nearest-neighbour
interaction on the staggered-flux phase in the Hubbard model. The
nearest-neighbour interaction has the form: $V\sum_{\langle ij
  \rangle}n_in_j$, where $\langle ij \rangle$ denotes the
nearest-neghbour pairs. This interaction can be added  to the
Hamiltonian of Eq. (\ref{Hubb}).
 In the following, we choose to look
at the point with a fixed on-site interaction $U=3.2$. In
Fig. \ref{delta_v}(a), the grand potential difference
$\Delta\Omega=\Omega(\Delta t, \Delta\phi)-\Omega(\Delta t)$ is
plotted as a function of nearest-neighbour interaction $V$ at
half-filling. For a positive $V$,  $\Delta\Omega$ is increasing to
zero as $V$ increases, implying that a positive repulsive $V$
disfavors the staggered-flux phase. However, adding a negative $V$
will decrease the value of $\Delta\Omega$. This means that an
attractive nearest-neighbour interaction $V$ is favorable for the formation
of the staggered-flux phase. However, for negative $V$ if  $|V|$ is large
enough  (e.g. $V<-0.5$), there is no stable VCA solution any more. The
reason might be that
a CDW order will form for very large attractive interaction and the
staggered-flux phase is not stable any more.

The effect of nearest-neighbour interaction $V$ on system away from
half-filling is shown in Fig. \ref{delta_v}(b) and (c) for $V=0.5$ and
$V=-0.5$ respectively, where the $\Delta\Omega=\Omega(\Delta\mu,\Delta
t, \Delta\phi)-\Omega(\Delta\mu,\Delta t)$ is plotted as a function of
chemical potential $\mu$. For positive $V=0.5$,  $\Delta\Omega$
increases slightly when $\mu$ deviates from the half-filling point
$\mu=3.6$ but is still inside the Mott gap region (i.e,
$3.5<\mu<3.7$).  $\Delta\Omega$ increases rapidly to zero as the
doping (both particle and hole) becomes nonzero ($\mu<3.5$ or
$\mu>3.7$). For negative $V=-0.5$, $\Delta\Omega$ remains nearly
constant in the Mott gap region (i.e, $-0.425<\mu<-0.375$). However,
interestingly, it decreases as the system is dopped (i.e, $\mu<-0.425$
or $\mu>-0.375$). Note that in both Fig. \ref{delta_v}(b) and (c), the
chemical potential corresponds to a 
density  between 0.9 and 1.1. Here, we do not investigate the large doping
region, in which a change of the particle sector in VCA is
needed. Athough the grand potential difference of staggered-flux phase
in Fig. \ref{delta_v}(c) is decreasing with doping, it is still hard
to compete with the Antiferromagnetic phase. For example, the grand
potential difference of Antiferromagnetic phase,
$\Omega(\Delta\mu,\Delta t, h_{AF})-\Omega(\Delta\mu,\Delta t)$, is
$-0.0317$, which is much lower than that of staggered-flux phase
($-0.008$) at the same point.

Therefore, our calculation shows that a negative nearest neighbour
interaction and finite doping is energetically favorable for the
formation of staggered-flux phase.

\subsection{Staggered flux
phase in the three band
  Hubbard model}
\label{3band}

Our work can be naturally extended to consider the three band Hubbard model 
to search for the existence of the spontaneous time-reversal symmetry breaking 
states. Such states are still 
under debate and 
have probably been observed in true materials~\cite{ka.ro.02,li.ba.08}.
In the prototype realsitic systems, the high-Tc 
cuprates, a crucial role is played by the CuO$_2$ planes, in which the oxygen orbitals 
are explicit degrees of freedom.
The electron dynamics in CuO$_2$ plane can be described by a minimal
model, the three band Hubbard model, containing the copper
$d_{x^2-y^2}$ orbital and oxygen $p_x$ and $p_y$ orbitals.
Experiments show evidence for time-reversal symmetry breaking in BSCCO
with photoemmision \cite{ka.ro.02} and in HgBa$_2$CuO$_{4+\delta}$
with polarized neutron diffraction \cite{li.ba.08}.  Early theoretical
works discussed the flux phase in the single band model with current order and
proposed it in conection to the pseudo-gap state in high-Tc cuprates
\cite{AffleckPRB88,MarstonPRB89,ch.la.01}.
In order to stabilize the flux phase, Varma shows in a mean-field
approach for three band model the necessity to go beyond on-site
interactions for an explicit treatment of the nearest-neighbor
interactions between copper and oxygen
~\cite{varm.97.nf,varm.99,si.va.02,varm.06}. 
Orbital currents were also obtained by more accurate theoretical
calculations for multi-band Hubbard models with or without the
inclusion of  oxygen orbitals~\cite{ch.ga.07,we.la.09}.
Within this subsection we show some preliminary results of VCA
calculations for the time-reversal symmetry breaking phase in three
band Hubbard model.

The three band Hubbard model under consideration is described by the Hamiltonian: 
\begin{eqnarray}
H &=& \sum_{<i,j>\sigma} t_{dp}^{ij} ( d_{i\sigma}^\dagger p_{j\sigma} + h.c. ) +
      \sum_{<j,k>\sigma} t_{pp}^{jk} ( p_{j\sigma}^\dagger p_{k\sigma} + h.c. ) \nonumber\\
   && + U_d \sum_{i} n_{i\uparrow}^d n_{i\downarrow}^d 
      + U_p \sum_{j} n_{j\uparrow}^p n_{j\downarrow}^p
      + V_{dp} \sum_{<i,j>\sigma} n_{i}^d n_{j}^p \nonumber\\
   && + (\epsilon_d-\mu) \sum_{i\sigma} n_{i\sigma}^d 
      + (\epsilon_p-\mu) \sum_{j\sigma} n_{j\sigma}^p 
\end{eqnarray}
Here, $d_{i\sigma}^\dagger$ and $p_{j\sigma}^\dagger$ create a hole in
copper $3d_{x^2-y^2}$ and oxygen $2p_\delta$ ($\delta=x,y$) orbitals;
$\epsilon_d$ and $\epsilon_p$ are the on-site energy for Copper and
Oxygen orbitals; $\mu$ is the chemical potential; $t_{dp}^{ij}$ and
$t_{pp}^{jk}$ are the neighbouring Cu-O and O-O hopping
amplitude. $U_d$, $U_p$, and $V_{dp}$ are the Coulomb repusion when
two holes sit on the same Copper orbital, the same Oxygen orbital,
or on the neighbouring Cu and O orbitals. 
One difficulty associated with the study of the CuO$_2$ planes is
the large number of parameters, giving rise to a huge phase space.
Therefore, the choice of parameters for the three band model of copper
oxides has been extensively discussed in the literature
\cite{hy.sc.89,mc.an.90,do.mu.92}.
Our choice corresponds to the following realistic value of parameters
$t_{pp}^{jk}=0.5$, $\Delta=\epsilon_p-\epsilon_d=3.0$, $U_d=8.0$,
$U_p=3.0$, and $V_{dp}=0.5$,
where  we have taken the Cu-O hopping $t_{dp}^{ij}$ as unit of the energies.

\begin{figure}[h]
\includegraphics[width=0.9\columnwidth]{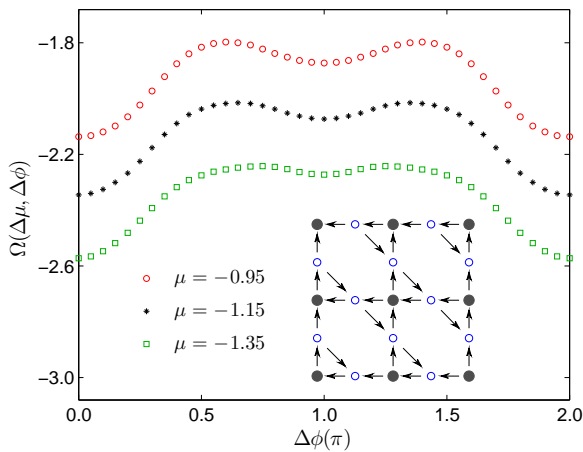}
\caption{(Color online). Grand potential
  $\Omega(\Delta\mu,\Delta\phi)$ of the three band Hubbard model as a
  function of the values of variational parameter $\Delta\phi$. The
  inset shows the phase configuration used in the calculation.
}
\label{omega_3band}
\end{figure}

In the inset of Fig. \ref{omega_3band} we show the CuO$_2$ plane in which 
a $2 \times 2$ unit cell plaquette is taken as the reference system for the VCA
calculations. Thus, a plaquette contains four 
unit cells each of them
 being composed of a Cu ($d_{x^2-y^2}$) and two O ($p_x$, $p_y$) sites. 
The implementation of time-reversal symmetry breaking 
within the three band model follows the description given in 
Sec. \ref{vca}. In the present calculations we have used a reduced 
number of variational parameters, namely the shift in the chemical 
potential $\Delta\mu$ as well as the shift $\Delta\phi$ of the phase 
on each hopping terms, the later having the form:
\begin{eqnarray}
 H_{TR}'&=& \sum_{<i,j>\sigma} t_{dp}^{ij} e^{i\Delta\phi_{dp_\delta}} ( d_{i\sigma}^\dagger p_{j\sigma} + h.c. ) \nonumber \\
        &+& \sum_{<j,k>\sigma} t_{pp}^{jk} e^{i\Delta\phi_{pp}} ( p_{j\sigma}^\dagger p_{k\sigma} + h.c. ) 
\end{eqnarray}
$\Delta\phi_{dp_\delta}$ ($\delta=x,y$) shifts the phases on the bonds 
between neighbouring Cu-O atoms 
inside the $2 \times 2$ reference plaquette.
Similarly, $\Delta\phi_{pp}$ shifts the phases on the bonds between neighbouring
O-O atoms
whithin the same 
plaquette. Note that the magnitude of the shifts in the phases 
are not necessarily equal. In fact, we have performed the computations 
with four variational parameters: 
$\Delta\phi_{dp_x}$, $\Delta\phi_{dp_y}$,
$\Delta\phi_{pp}$ and $\Delta\mu$. The 
direction and distribution of the phases on each hopping bonds
are taken according to the ``current pattern'' 
proposed by Varma, Ref.~\cite{varm.06}. According to our calculations, 
none of the two current (phase) patterns considered in Varma's mean-field
proposal \cite{si.va.02,varm.06} are stable. Whithin the calculations with 
four variational parameters the saddle piont of grand potential $\Omega$ 
always locates at $\Delta\phi_{pp}=\Delta\phi_{dp_x}=\Delta\phi_{dp_y}=0$.
Therefore no spontaneous time-reversal symmetry breaking phases are found in our
calculations.

In Fig. \ref{omega_3band} we plot the grand potential $\Omega(\Delta\mu,\Delta\phi)$ 
as a function of $\Delta\phi$  for different values of $\mu$.
For the sake of illustration, 
the calculation 
has been performed using the simplifying assumption of equal phase shifts, i.e. 
when the values of $\Delta\phi_{pp}$, $\Delta\phi_{dpy}$, and
$\Delta\phi_{dpy}$ are all equal. 
The phase (current) pattern is also shown in the inset of figure Fig. ~\ref{omega_3band}.
This is the pattern denoted as the type $\Theta_{II}$ in the orginal 
work of Varma~Ref. \cite{varm.06}. The saddle point in the grand potential $\Omega$ is 
obtained for zero phase on bonds, and turnning on phase around zero will lead to an
increase of $\Omega$. The situation when all three $\Delta\phi$ are different is similar.

Note that additional variational parameters such as $\Delta t$ are required 
to complete our study on three band Hubbard model, however this is out  
of the scope of present paper.

\section{Conclusion}
\label{con}

The central aim of this paper was to 
extended the VCA method to study the time-reversal
symmetry broken phase. We proposed to use two variational parameters,
the hopping $\Delta t$ and phase factor $\Delta\phi$, to study the
staggered flux phase in the 2D Hubbard model. 
Our calculation suggest a metastable staggered flux phase in 
 the intermediate and small $U$
regions. 
For the case of half filling,
the flux phase appears to be the most stable at
 $U \approx 3.2$.
This implies that the staggered flux phase, or at least its
fluctuations,  
are more likely to be observed
 in the small $U$  than in large $U$ region.
We also studied
the stability of the staggered flux phase by comparing its grand potential
with those of the superconductor and antiferromagnetic phases. We
found that one of these latter phases is always more stable than the
 staggered flux phase both 
at as well as  away from half filling. 

From our results on the one-band 2D Hubabrd model we cannot draw a
clear picture of which interactions are crucial to cause the orbital
currents, although we clearly observe that a nearest-neighbour
repulsion ($V>0$) decreases the current, that would be present in the
absence of the intersite interaction. In contrast, an attractive
nearest-neighbour interaction ($V<0$) leads to an increase in the
current comparing with the $V=0$ case. The staggered-flux order that
emerges in the presence of $V<0$ was previously reported within the
strong-coupling regime \cite{StanescuPRB01}. Our numerical results
demonstrate that such result is also applicable to the intermediate
values of $U$.  At last, we also extented the method for time-reversal
symmetry breaking phase to study the three band Hubbard model. By
using 
only the
shifts of phases on each hopping terms as variational parameter, we
did not find a stable VCA saddle point for the staggered flux
 phase.

\begin{acknowledgments}
L.C. acknowledges discussions with P. Chakraborty and
J. Kunes. X.L. thanks H. Allmaier, Z. B. Huang, and S. P. Kou for helpful discussions.
This work was supported by the National Natural Science Foundation
of China (No. 11004164 and No. 10974163) and by the Austrian Science Fund
(FWF) P18551-N16.
\end{acknowledgments}

%\bibliography{/home/xiancong/work/documents/bibliography/references_database,/home/xiancong/work/documents/bibliography/cold_atom}

\begin{thebibliography}{49}
\bibitem[{foo()}]{footnote}
\bibinfo{note}{Due to the artificial symmetry breaking caused by the cluster
  partition, the order parameter is not uniform. Therefore, we obtain it as an
  average of the modulus of the current over all bonds \textit{inside} the
  reference cluster.}
\end{thebibliography}

\end{document}